\documentclass[pra,aps,showpacs,twocolumn,floatfix]{revtex4}
\usepackage{epsfig} \usepackage{graphics} \usepackage{bm} \usepackage{amssymb} \usepackage{color}

\begin{document}


\title{Stability of the density-wave state of a dipolar condensate in a pancake trap}
\author{O.~Dutta\footnote{Corresponding author.
Electronic address: dutta@physics.arizona.edu}}
\affiliation{B2 Institute, Department of Physics and College of Optical Sciences,
The University of Arizona, Tucson, AZ 85721, USA}
\author{R.~Kanamoto\footnote{Present Address: Ochanomizu University, Tokyo 112-8610, Japan}}
\affiliation{B2 Institute, Department of Physics and College of Optical Sciences,
The University of Arizona, Tucson, AZ 85721, USA}
\author{P.~Meystre}
\affiliation{B2 Institute, Department of Physics and College of
Optical Sciences, The University of Arizona, Tucson, AZ 85721, USA}
\date{\today}


\begin{abstract}
We study a dipolar boson-fermion mixture in a pancake geometry at absolute zero temperature, generalizing our previous work on the stability of polar condensates and the formation of a density-wave state in cylindrical traps.
After examining the dependence of the polar condensate stability on
the strength of the fermion-induced interaction, we determine the transition
point from a ground-state Gaussian to a hexagonal density-wave state.
We use a variational principle to analyze the stability properties of those density-wave state.
\end{abstract}
\pacs{03.75.-b, 42.50.-p, 33.80.-b, 03.65.-w} \maketitle


\section{Introduction}

The recent realization of a Bose-Einstein condensate of chromium
atoms \cite{chbec1, chbec2} opens up the study of
quantum-degenerate gases that interact via the long range,
anisotropic magnetic dipole interaction. This is an anisotropic
and long-range interaction that leads to the appearance of a
wealth of new properties past those characteristic of systems with
isotropic interactions~\cite{oplat}. In the case of $^{52}$Cr
$s$-wave interactions are normally much stronger than the dipolar
part, but it has been shown experimentally that by applying a
magnetic field the $s$-wave scattering length can be lowered to
values comparable to the Bohr radius \cite{pfau1}, so that the
dipole interaction dominates the system. In the regime where the
dipole-dipole interaction is dominant the condensate is
characterized by the existence of a metastable state whose
properties depends on the trapping geometry \cite{dpstab, you,
odell,goral,bohn,dutta}, as was experimentally demonstrated in
Ref.~\cite{pfau2}. A roton feature has also been predicted to
exist in these systems for appropriate parameters. There would
considerable interest indeed in accessing the associated roton
instability \cite{rotmax, coop, Uf}, as it is characterized by the
spontaneous generation of a periodic density modulation in the
condensate \cite{rica}. Unfortunately that state is unstable
against collapse, but we found recently that this difficulty can be
circumvented by the addition of a small fraction of
non-interacting fermions, resulting in a significant stabilization of
the bosonic system in cylindrical traps
\cite{dutta1}. The simultaneous trapping of bosonic and fermionic
isotopes of chromium \cite{gor} also points to an interesting
directions, with the possibility to observe novel ground and
metastable phases in quantum degenerate polar boson-fermion
mixtures.

This paper extends our previous study of dipolar boson-fermion
mixtures from cylindrical traps to pancake geometries, showing that dipolar
bosons can be stabilized considerably by increasing the boson-fermion
$s$-wave scattering length, and more importantly the density-wave states of
dipolar bosons can be stable for typical, zero-field boson-fermion interaction
strengths. Section~II describes the induced potential created by fermions on bosons,
and Sec.~III discusses the general properties of
the energy functional as a function of boson-fermion interaction
using a variational Gaussian ansatz. Section IV presents an analysis
of the transition point to the density-wave state for various
combinations of trap aspect ratios and boson-boson contact interaction
strengths. Section V introduces a density-wave ansatz in the form of a
series of shifted Gaussians. The stability of these density-wave states
is determined by varying the width of each Gaussian to seek minima in the
interaction energy for various boson-boson $s$-wave interaction. We find a
condition for the stability of these types of density modulated
states. Finally, Section VI is a summary and outlook.

\section{Fermion-induced interaction in polar condensates}

We consider a mixture of $N_b$ dipolar bosons of mass $m_b$ and $N_f$ single-component
fermions of mass $m_f$ confined in a pancake-shaped trap
characterized by a tight harmonic potential of frequency $\omega_{z}$ along
the $z$-axis and a softer harmonic potential of frequency $\omega_{\bot}$
in the transverse direction. A polarizing external electric or magnetic field,
is taken to be along the $z$-axis . The dipole-dipole interaction between two
bosonic particles separated by a distance $r$ is then
\begin{equation}
V_{\rm dd}(r) = g_{\rm dd}
{\left ( 1 - \frac{3z^2}{r^2} \right )}\frac{1}{r^3},
\end{equation}
where $g_{\rm dd}$ is the dipole-dipole interaction strength. In the
mean-field approximation, the energy functional for the order
parameter $\phi(\bm{r})$ of the dipolar condensate can be expressed as
\begin{eqnarray}\label{energy}
  E\!\!&=&\!\!\!\int \phi^{*}(\bm{r}) H_0 \phi(\bm{r}) d^3r + \frac{g N_b}{2}
  \int | \phi(\bm{r}) |^4 d^3r \\
  &+&\!\!\frac{N_b}{2} \! \int\!\!\!\!\int \! | \phi(\bm{r})  |^2
  V_{\rm dd}(\bm{r} - \bm{r}') | \phi(\bm{r}') |^2 d^3r d^3r' \!+\!
  E_{\rm ind}, \nonumber
\end{eqnarray}
where
\begin{equation}
H_0 = -\frac{\hbar^2}{2m_b}\nabla^2  +\frac{ m_b \omega^2_{z}}{2} \left
[\lambda^2 \right ( x^2 + y^2 \left ) + z^2 \right ]
\end{equation}
is the sum of the kinetic energy and the trapping potential and $\lambda=
\omega_{\bot}/\omega_z$. The second term in the energy functional
(\ref{energy}) denotes the contact interaction between bosons,
characterized by the strength $g=4 \pi \hbar^2 a_{bb}/m_b$ with $a_{bb}$
being the $s$-wave scattering length, and the third term describes the nonlocal dipole-dipole interaction between bosons. Finally, the last term $E_{\rm ind}$ accounts for the fermion-induced interaction $V_{\rm ind}(\bm{k})$ between bosons, given in linear response theory~\cite{np} as follows.

We assume a contact boson-fermion interaction of strength $g_{bf}=2\pi\hbar^2 a_{bf}/m_r$, where $a_{bf}$ is the boson-fermion $s$-wave scattering length and $m_r=m_bm_f/(m_b+m_f)$ is the reduced mass. The boson-fermion interaction energy has the form $g_{bf} \int n_f(\bm{k}) n(-\bm{k})d^3k$, with $n(\bm{k})$ and $n_f(\bm{k})$ being the bosonic and fermion densities in the momentum space. The linear response of the fermions to a bosonic density
$n(\bm{k})$ can be expressed as $n_f(\bm{k})= V_{\rm ind}(\bm{k})n(\bm{k})$, so that for the effect of the fermions on the bosonic energy functional is
\begin{equation}\label{fermi}
E_{\rm ind}=\frac{1}{2} \frac{g_{bf} N_b}{(2\pi)^3}
\int V_{\rm ind}(\bm{k}) n(\bm{k}) n(-\bm{k}) d^3k.
\end{equation}
The explicit form of the induced potential is
\begin{equation}
V_{\rm ind}(\bm{k})= g_{bf}\chi_f(\bm{k}),
\end{equation}
where $\chi_f$ is the density response function~\cite{np}. It is related to the dynamical structure factor $S(\bm{k},\omega)$, which is the probability of exciting particle-hole pairs with momentum $\bm{k}$ out of the Fermi sea,
by $\chi_f(\bm{k})=-2 \int^{\infty}_0 d\omega'[S(\bm{k},\omega')/\omega']$. For a non-interacting single-component Fermi system we have
\begin{equation}\label{struc}
S(\bm{k},\omega)= \mathop{\sum_{p<k_f}^{\infty}}_{|\bm{p}+\bm{k}|>k_f}
\delta(\omega-\omega^0_{\bm{p} \bm{k}}),
\end{equation}
where $p=|\bm{p}|$, $k_f$ is the Fermi momentum, the excitation
energy is $\omega^{0}_{\bm{p} \bm{k}}=pk \cos \theta/m_f
+k^2/(2m_f)$, and $\theta$ is the relative angle between $\bm{p}$
and $\bm{k}$. This expression assumes that the fermions are locally free, so that there is a local Fermi sphere in momentum space.
In this local density approximation, $k_f$ is given by \cite{peth}
\begin{equation}\label{kf}
  k_f \ell_z= 1.9 N^{1/6}_f \lambda^{1/3} \sqrt{\frac{m_f}{m_b}},
\end{equation}
where $\ell_{z}=\sqrt{\hbar/(m_b \omega_{z})}$ is the oscillator length in the $z$ direction.

Following Ref.~\cite{dutta1} the induced potential $V_{\rm ind}(\bm{k})$ is given by
\begin{eqnarray}\label{indu}
V_{\rm ind}(k) \!\!=\!\!
\left\{
\begin{array}{lll}
\displaystyle{\!\!g_{bf}\nu\left [ -1 + \sum_{n=1}^\infty
\left(\! \frac{k}{2k_f} \!\right)^{\!\! 2n}\!\!\!\frac{1}{4n^2-1}  \right]},
&&\!\!\! k < 2k_f,\\
\displaystyle \!\!- { {g_{bf} \nu} \sum_{n=1}^\infty \left(\! \frac{2k_f}{k}\! \right)^{\!\! 2n}\!\!\!\frac{1}{4n^2-1} },
&& \!\!\! k > 2k_f,
\end{array}
\right.
\end{eqnarray}
where $\nu=k_f m_f/(\pi \hbar)^2$ is the three-dimensional fermionic density
of states, and $k^2=k^2_x+k^2_y+k^2_z$ is the square of the fermionic momentum. The induced potential is attractive for very low momenta and goes to zero with increasing momenta.

Throughout this paper, we consider the parameters corresponding to a
$^{52}$Cr-$^{53}$Cr mixture, and fix the number of fermions to be $N_f=10^{3}$
unless otherwise stated. The zero-field bosonic $s$-wave interaction
for $^{52}$Cr is taken to be $a_{bb}=103a_0$, and $a_{bf}=70a_0$ for the
boson-fermion scattering length \cite{tpfau}.

\section{ Stability of the Gaussian dipolar condensate}

We now proceed to determine the stability of the dipolar condensate,
using a variational ansatz in the parameter space of the fermion-induced
interaction and dipolar strength. The variational wave function is taken as the Gaussian
\begin{equation}\label{ph}
\phi (\bm{r}) = \frac{1}{\sqrt{\pi^{3/2} d^2 d_z}} \exp \left [ - \frac{x^2 + y^2}{2 d^2}
- \frac{z^2}{2 d^2_z}\right ],
\end{equation}
with $d, d_z$ being the variational parameters and $ \int d^3
r |\phi(\bm{r})|^2=1$.
Substituting this Gaussian ansatz and its Fourier transform into
Eqs.~(\ref{energy}) and (\ref{fermi}) yields the energy $E_g$ of the condensate as
\begin{widetext}
\begin{eqnarray}\label{enfun}
\frac{ m_b \ell^2_{z}}{\hbar^2} E_g\!\!\!&=&\!\!\!
\frac{1}{2} \left ( \frac{1}{2}+\eta^2 \right )
\left(\frac{\ell_z}{d_z}\right)^2 +
\frac{1}{2} \left ( \frac{1}{2} + \frac{\lambda^2}{\eta^2} \right )
\left(\frac{d_z}{\ell_z}\right)^2 \nonumber\\
&+&  {g_{\rm 3d}}\left(\frac{\ell_z}{d_z}\right)^3
\left [ \frac{2}{3} \eta^2 - F(\eta^{-1}) +
\left(\frac{d_z}{\ell_z}\right)^3 \left\{g^{<}E^{<}_1+
g^{>}E^{>}_1+g_{\rm ind}  (E^{<}_2 -E^{>}_2+E^{<}_3)\right\}
\right ],
\end{eqnarray}
\end{widetext}
where $\eta=d_z/d$, and $\tilde{k}_f=k_f \ell_{z}$.
The derivation of algebraic forms of the interaction-energy terms
$E_1^{<}, E_1^{>}, E_2^{<}, E_2^{>}, E_3^{<}$ and of
the function $F$ are straightforward but lengthy, and their derivations
are relegated to an Appendix. In Eq.~(\ref{enfun}) we have also introduced the
effective three-dimensional dipole-dipole interaction
$$
g_{\rm 3d}=\frac{m_b N_b g_{\rm dd} }{\sqrt{2 \pi} \hbar^2 \ell_{z}}.
$$
The coefficients $g^<$ and $g^>$ find their origin in the momentum-independent contact interaction, given by the summation of the $s$-wave boson-boson scattering and the constant terms in the induced interaction~(\ref{indu}). They are given explicitly by
\begin{eqnarray}\label{contact}
g^{<} &=& \frac{g-g_{bf}^2\nu}{4\pi g_{\rm dd}}, \nonumber\\
g^{>} &=& \frac{g}{4\pi g_{\rm dd}}.
\end{eqnarray}
Finally
$$
g_{\rm ind}=\frac{g^2_{bf} \nu} {48 \pi g_{\rm dd}\tilde{k}^2_f}
$$
is due to the nonlocal induced interaction.
We remark that the effective contact interaction consists of the boson-boson contact interaction as well as the momentum-independent part of the dipole-dipole interaction and of the induced interaction. From Eqs.~(\ref{enfun}) and (\ref{contact}) we have that for low momenta
\begin{eqnarray}\label{gs}
g_s=\frac{2}{3}+g^{<}.
\end{eqnarray}

It is known that in boson-fermion mixtures without dipolar interaction, phase separation occurs
when $g_s$ becomes negative \cite{cohen,Tosi, bsfer}. In this paper, in contrast, we restrict
our considerations to the case $g_s>0$ by changing the boson-boson contact interaction $g$,
so that phase separation does not take place.

\begin{figure}[ht]
\begin{center}
\epsfig{file=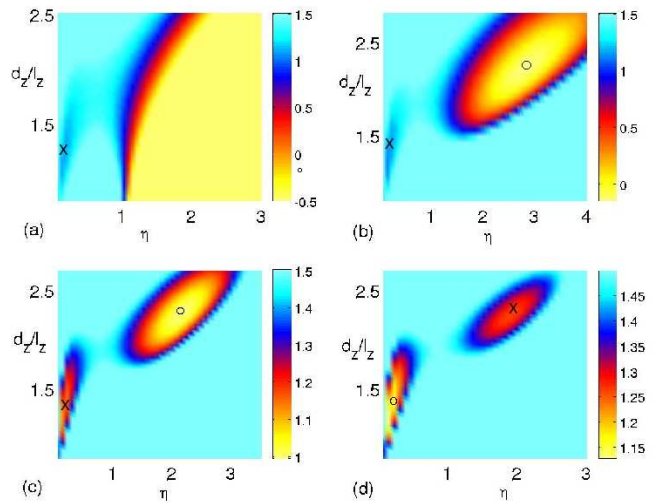,width=9.5cm}
\caption{\label{fig1}
Energy landscape $E_g$ obtained by the variational Gaussian ansatz Eq.~(\ref{ph})
as a function of $\eta=d_z/d$ and $d_z/\ell_z$ at a constant dipolar interaction strength $g_{\rm 3d}=30$.
In these figures the energy is cut off at $E_g=1.5$ from above
and $E_g=-0.5$ from below for viewability.
(a) Behavior of $E_g$  in the absence of
the fermion-induced interaction $g_{\rm ind}=0$
(b) Energy $E_g$ for $g_{\rm ind}=0.02$.
The plot is characterized by presence of two minima:
(i) the true, narrow Gaussian ground state located at
$(\eta, d_z/\ell_z)\simeq (3, 2.3)$; and (ii) a pancake-shaped metastable state
located at $(\eta, d_z/\ell_z)\simeq (0.2, 1.4)$.
(c) Energy landscape for $g_{\rm ind}=0.06$. The energies of the ground- and
metastable states approach each other.
(d) The broad Gaussian metastable state at $(\eta, d_z/\ell_z)\simeq (0.2,1.4)$
eventually becomes the true ground state for $g_{\rm ind}=0.07$.
In the figures the global (ground state) and local minima (metastable state)
are indicated by a circle and cross, respectively.}
\end{center}
\end{figure}

The energy functional~(\ref{enfun}) was minimized with respect to
$\eta$ and $d_z/\ell_z$ for various parameter values, with our results summarized in Fig.~\ref{fig1}. Without the fermion induced interaction, $g_{\rm ind}=0$, the ground-state energy of the system is not bounded from below, and
the strong trapping in $z$ direction creates a {\it local minimum} in the
energy landscape as a function of $d_z$ and $\eta$ [Fig.1 (a)].
For finite induced interactions $g_{\rm ind} > 0$, we find in contrast that the energy landscape is characterized by two minima, as shown in Fig.~1 (b), (c), and (d). In Fig.~1(b) the global minimum, which occurs for $(\eta,d_z/\ell_z) \simeq (3, 2.3)$, is a Gaussian state with narrow width in the transverse $x$-$y$ plane. The additional local minimum close to $(\eta, d_z/\ell_z)\simeq (0.2, 1.4)$ is a metastable state.

With increasing boson-fermion interaction $g_{\rm ind}$, though, the ground-state energy corresponding to the global minimum at $\eta>1$ approaches that of the local minimum [Fig. 1 (c)], and
these minima eventually reach equal energies at a critical value of $g_{\rm ind}$. The new ground state past that point is characterized by the parameters $\eta<1$ and
$d_z \sim \ell_z$ [Fig.~1 (d)], that is, it is a wide
Gaussian in $x$-$y$ plane.

Experimentally one may either vary the boson-boson
$s$-wave scattering length while keeping the boson-fermion scattering length constant,
or vary $g_{bf}$ with constant $s$-wave interaction $g$.
In a boson-fermion mixture of chromium isotopes,
typically $g_{\rm ind} \sim 10^{-3}$ is small,
which corresponds to an energy landscape that resembles that of Fig.~\ref{fig1}(b)
and the broad Gaussian corresponds to a metastable state. Such a metastable state has
been achieved experimentally in experiments by the Stuttgart group \cite{pfau2}. A system that offers
the potential to reach the regime of Fig.~\ref{fig1}(d) is provided by a mixture of bosonic ${}^{87}$Rb
and fermionic ${}^{40}$K with the scattering length of rubidium atoms tuned close to zero. In that mixture,
a zero-field scattering length
$a_{bf} \approx 250a_0$ \cite{gold} gives $g_{\rm ind} \sim 0.2$.


\section{Transition to a density wave}

The excitation spectrum of condensates dominated by a dipolar interaction is
predicted to exhibit a roton minimum \cite{rotmax, Uf}.
As a consequence, a bosonic density-modulated state in the $x$-$y$ plane
may arise as a local minimum, and it may actually have a lower energy than the
metastable Gaussian state and for high enough bosonic densities \cite{rica}. This section discusses
the transition to the appearance of such a state as a function of number of
bosonic particles for various combinations of trap ratio.

We proceed by introducing the new variational wave function that describes a density-wave structure with triangular symmetry,
\begin{widetext}
\begin{equation}\label{hex}
\phi^{\rm t}_{\rm dw}(x,y,z) = \phi(x,y,z)
\left [ a_0+ \sum_{n=1}^{\infty} a_n \left \{ \cos\!\left(\frac{n \tilde{k}_0 x}{d_z}\right) +  2\cos\!\left(\frac{n\tilde{k}_0 x}{2 d_z}\right)
\cos\!\left( n \frac{ \sqrt{3} \tilde{k}_0 y}{ 2 d_z}\right) \right \}  \right ],
\end{equation}
where $\phi(x,y,z)$ is defined in Eq.~(\ref{ph}), $n$ is an integer, and
$\tilde{k}_0 = k_0d_z \ll \eta=d_z/d$.
Substituting this trial wave function~(\ref{hex}) into Eq.~(1), we find that the scaled excess energy of the density-modulated state relative to the Gaussian state
$$
\epsilon(\tilde{k}_0, a_1,a_2,a_3)= \frac{2m d^2_z}{\hbar^2}(E_{\rm dw} - E_{g})
$$
is approximately given by
\begin{eqnarray}\label{hexen}
\frac{\epsilon(\tilde{k}_0, a_1,a_2,a_3)}{3}
&\approx&    \frac{\tilde{k}^2_0}{4} (a^2_1+4a^2_2+9a^2_3)
+ \frac{g_{\rm 3d}n_d}{4}
\left[ (a^2_1+2a_0a_1+a_1a_2+a_2a_3)^2 V_{\rm eff}(\tilde{k}_0)
\right.\nonumber\\
&+& (a^2_1+2a_1a_2)^2 V_{\rm eff}(\sqrt{3}\tilde{k}_0)
+\frac{1}{4}(a^2_1+2a^2_2+4a_0a_2+2a_1a_3)^2
V_{\rm eff}(2\tilde{k}_0) \nonumber\\
&+&\left. {2} (a^2_1a^2_2+a^2_1a^2_3+a^2_2a^2_3)
V_{\rm eff}(\sqrt{7}\tilde{k}_0)
+  (2a_0a_3+a^2_3)^2 V_{\rm eff}(3\tilde{k}_0) \right],
\end{eqnarray}
where $n_d\equiv \eta^2 \ell_z/d_z$~\cite{nd}.
The normalization condition reads $a^2_0+3(a^2_1+a^2_2+a^2_3)/2=1$,
and the effective transverse potential is
\begin{eqnarray}\label{veff}
V_{\rm eff}(\tilde{k}_0) \approx
\left\{
\begin{array}{lll}
\displaystyle{\frac{2}{3} + g^{<} - \sqrt{\frac{\pi}{2}}\
\tilde{k}_0\  {\rm erfcx} \left ( \frac{\tilde{k}_0}{\sqrt{2}} \right )
+ {g_{\rm ind}}\left(\frac{\ell_{z}}{d_z}\right)^2 \tilde{k}^2_0},
&& \tilde{k}_0 < 2k_fd_z,\\
\displaystyle  {\frac{2}{3} + \frac{g}{4\pi g_{\rm dd}}
- \sqrt{\frac{\pi}{2}}\ \tilde{k}_0\ {\rm erfcx} \left ( \frac{\tilde{k}_0}{\sqrt{2}} \right )
- {g_{\rm ind}}\left(\frac{\ell_{z}}{d_z}\right)^2 \frac{(2k_fd_z)^4}{\tilde{k}^2_0}
},
&& \tilde{k}_0 > 2k_fd_z,
\end{array}
\right.
\end{eqnarray}
\end{widetext}
In evaluating Eq.~(\ref{hexen}) we kept only the first four terms $n=0,\dots, 3$
of Eq.~(\ref{hex}) as the energy converges at the transition point.
The magnitude of the error in that approximate expression is estimated to be of the order of $e=\exp(-\tilde{k}^2_0\eta^2/4)$. In the subsequent calculations we consider values of $\tilde{k}_0$ such that this error is less than or
on the order of $10^{-6}$.

\begin{figure}[h]
\begin{center}
\epsfig{file=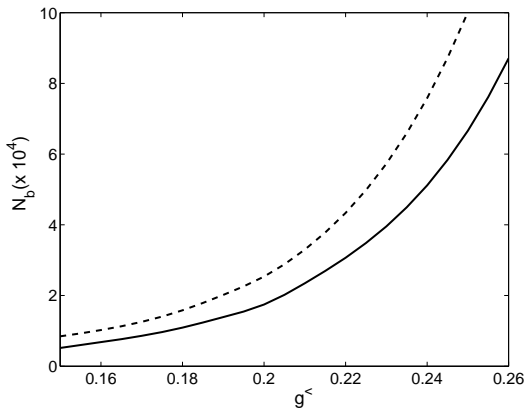,width=7cm}
\caption{\label{fig2}
Critical number of bosonic ${}^{52}$Cr
atoms needed for the transition from the
Gaussian state to a hexagonal density-wave state for the trap aspect ratios
$\lambda=0.25$ (solid curve) and
$\lambda=0.22$ (dashed curve).
Here $\ell_z=0.15\mu$m.}
\end{center}
\end{figure}

We numerically determine the variational parameters that minimize
the excess energy $\epsilon(\tilde{k}_0, a_1,a_2,a_3)$ as a function of
$g^<$ for fixed $n_d$ and $g_{\rm 3d}$.
These results are summarized in Fig.~\ref{fig2}, which shows the critical
number of bosonic particles $N_b$ such that $\epsilon$ has a minimum for $\tilde{k}_0 \ne 0$.
By tuning the trap aspect ratio $\lambda=\omega_{\perp}/\omega_z$ to
higher values, the critical number of bosons $N_b$ required to achieve $\epsilon < 0$, that is, a
transition from a Gaussian to a density-wave metastable state with lower energy, is lowered
as a result of an increased $n_d$. In addition, we note that
the value of $\tilde{k}_0$ determined by the present
variational calculation matches closely the position of roton minima at
the excitation spectrum of the condensate
as in Ref.~\cite{rica}. We also checked the energy with density wave having
square symmetry and find that the triangular one always has lower energy.

\section{Density wave stability}

In the previous section we determined the transition to the density-wave instability, assuming the ansatz~(\ref{hex}). As already mentioned, this transition is related to the existence of a roton minimum in the excitation spectrum. However, assuming such a density wave does not provide an answer to the question of its stability against collapse. This is the issue that we address now, restricting our considerations to the regime where the boson-fermion interaction strength is close to its zero-field value $a_{bf}=70a_0$. This corresponds in the case of ${}^{52}$Cr-${}^{53}$Cr isotopes to $g_{\rm ind} \sim 10^{-3}$. This is the parameter regime characterized by a Gaussian metastable state, see Sec.~II and Fig.~\ref{fig1}(b).

Once the transition to the density-wave state has occurred
the condensate can also be described as a superposition
of shifted Gaussian wave functions within the two-dimensional $Z\equiv x+iy$ plane,
\begin{equation}\label{dw1}
\phi_{\rm dw}(Z, z)=\frac{1}{\sqrt{\mathcal{M} \pi^{3/2} \xi^2 d_z}} \sum_{j=1}^{\mathcal{M}}
\exp \left [ - \frac{(Z-Z_j)^2}{2 \xi^2} - \frac{z^2}{2 d^2_z} \right ]
\end{equation}
where $j$ is the index of the lattice site, $Z_j$ the lattice vectors generating periodic density modulations, $\xi$ the width of each density
peak, $\mathcal{M}$ the total number of density peaks, and $l$
the distance between neighboring density peaks. It is given by $l=2\pi/k_0$ where $k_0$ is the position of roton minimum in the excitation spectrum of the condensate.

Using the results of Sec.~III to investigate the stability of each Gaussian in Eq.~(\ref{dw1}), we now show that the presence of fermions
substantially stabilizes the density-wave state, noting that a stable density wave should be characterized by non-zero values of $\xi/d_z$ and $d_z/\ell_z$.

From Eq.~(\ref{dw1}), the total interaction energy is found to be ~\cite{sup}
\begin{widetext}
\begin{eqnarray} \label{enfun3}
\mathcal{E}_{\rm int} (\xi,d_z)&=&\frac{ m_b \ell^2_{z} E_{\rm int}}{\hbar^2}
\frac{\mathcal{M}}{g_{\rm 3d}} \nonumber\\
&=&
 \left(\frac{\ell_z}{d_z}\right)^3
\left [ \frac{2}{3} \left(\frac{d_z}{\xi}\right)^2
 - F(d_z/\xi)
+ \left(\frac{d_z}{\ell_z}\right)^3 \left\{g^{<}E^{<}_1
+ g^{>}E^{>}_1
+  g_{\rm ind}(E^{<}_2 - E^{>}_2 + E^{<}_3)\right\}
+ f(l, \xi, d_z) \right ], \nonumber\\
&&
\end{eqnarray}
\end{widetext}
where we have assumed that $\xi \ll l$, i.e., that neighboring Gaussian peaks
in Eq.~(\ref{dw1}) have little overlap. Here $E_1^{<}, E_1^{>}, E_2^{<}, E_2^{>},
E_3^{<}$ and $F$ have the same form as in the Appendix,
but $\sigma$ is now a function of $(\xi, d_z, \theta)$,
$\sigma=\sqrt{d^2_z \cos^2\theta+\xi ^2\sin^2\theta}$. The function $f(l, \xi, d_z)$
is responsible for the particular geometry of the density-modulated state. Other terms,
on the other hand, just arise from the energy of each individual Gaussian
and we can thus apply the results of Sec.~III to the present analysis.

Assuming a triangular crystal and including only the interaction between
nearest neighbors, we have
$$
f(l, \xi, d_z)=6 \int^{\infty}_0  V_{\rm eff}(\tilde{k})
\exp \! \left(
-\frac{\tilde{k}^2\xi^2}{2 d_z^2} \right)
J_0\!\left( \frac{\tilde{k} l}{d_z} \right) \tilde{k} d\tilde{k},
$$
with $\tilde{k}=kd_z$, the effective two-dimensional potential $V_{\rm eff}(\tilde{k})$
is defined in Eq.~(\ref{veff}), and $J_0$ is the zeroth-order Bessel function of the first kind.

\begin{figure}[ht]
\begin{center}
\epsfig{file=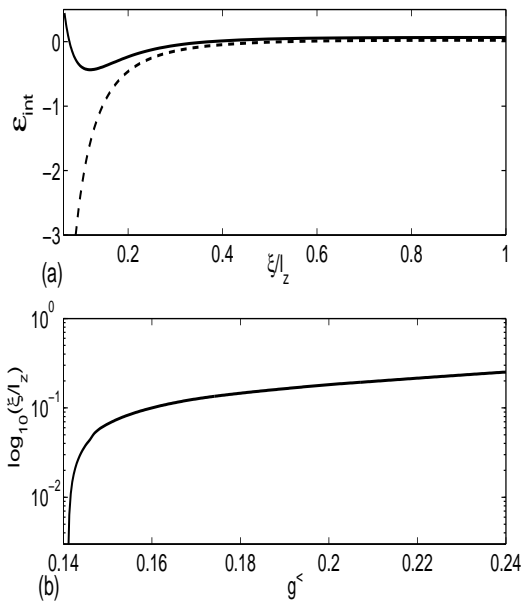,width=7cm,height=8cm}
\caption{\label{fig3}
(a) Interaction energy $\mathcal{E}_{\rm int}$  of Eq.~(\ref{enfun3}) as a function of
$\xi/\ell_z$ at $\lambda=0.25$, $\ell_z=0.15\mu$m, $d_z=1.5 \ell_z$, and $g^{<}=0.18$.
For $g_{bf}=0$ the energy is unbounded from below, but for $g_{bf}=70a_0$,
a minimum appears at $\xi/\ell_z\approx 0.1$.
(b) Log-scale plot of the normalized width $\xi/\ell_z$ that minimizes $\mathcal{E}_{\rm int}$ as a function of $g^{<}$.
}
\end{center}
\end{figure}

We now discuss the minimum of the energy functional $\mathcal{E}_{\rm int}(\xi, d_z)$ as a function of the effective strength of the contact interaction
$g_s$, which is varied by changing $g^<$, see Eq.~(\ref{gs}).
Without boson-fermion interaction, the interaction energy is a monotonically
increasing function of $\xi$ with $\mathcal{E}_{\rm int}(\xi \rightarrow 0, d_z) \rightarrow - \infty$.
As a result each gaussian of the density wave Eq.~(\ref{dw1}) collapses. A typical example of the
dependence of the interaction energy on $\xi$ is shown as the dotted curve in Fig.~3(a).
With non-zero boson-fermion interaction energy, in contrast, the interaction energy~(\ref{enfun3})
exhibits a minimum at a finite $\xi$ as illustrated by the solid curve in Fig.~3(a) f
or $d_z=1.8 \ell_z$ and $g^{<}=0.17$. The existence of that
minimum implies the stability of each Gaussian, that is, the stability of
the wave function~(\ref{dw1}).

\begin{figure}[h]
\begin{center}
\epsfig{file=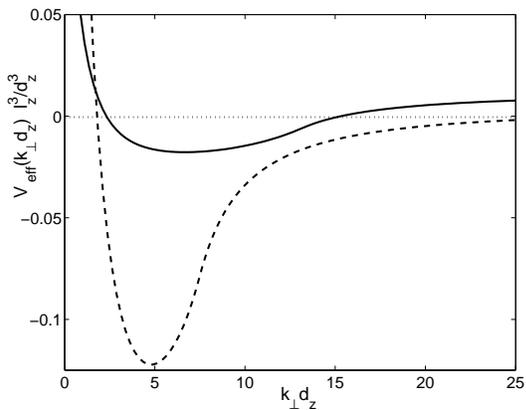,width=7cm}
\caption{\label{fig4}
Effective potential $V_{\rm eff}$ as a function of $k_{\perp}d_z$
for the parameters of Fig.~\ref{fig3}(b) for
$g^{<}=0.2$ (solid curve), and the critical value of $g^{<}\approx 0.14$ (dashed curve).}
\end{center}
\end{figure}

We can determine the dependence of the gaussian widths $\xi$ on
the $s$-wave boson-boson scattering length by changing $g^{<}$.
To do this we first minimize Eq.~(\ref{enfun3}) as a function of
$g^{<}$, treating $d_z$ and $ \xi$ as variational parameters.
Figure 3 (b) shows the resulting value $\xi/\ell_z$  as a function of
$g^{<}$ for $\lambda=0.25$, $N_f=1000$ and $\ell_z=0.15\mu$m. We observe that $\xi$ decreases
for smaller values of the contact interaction $g^{<}$, and for $g^{<} \lesssim 0.14$,
the minimum-energy state corresponds to the collapsed state, $\xi=0$. Below
that point the density wave is unstable.

This behavior can be understood from the effective potential by first integrating out
the $z$ direction to obtain the effective potential $(\ell_z/d_z)^3 V_{\rm eff}(k_{\perp}d_z)$,
where $V_{\rm eff}$ is given by Eq.~ (\ref{veff}) and $k_{\bot}^2=k^2_x+k^2_y$. The
width $d_z$ of the Gaussians is determined by the position of the minimum in
Eq.~(\ref{enfun3}). For $g^{<} \lesssim 0.14$, $V_{\rm eff}(k_{\bot}d_z)$ is
attractive for {\it all} high momenta, but at the critical value
$g^{<} \approx 0.14$, it approaches zero as $k_{\bot} \rightarrow \infty$, as
shown by the dashed curve in Fig.~\ref{fig4}. Hence the energy is always minimized
 when the wave function has zero width in the $x$-$y$ plane, i.e., $\xi \rightarrow 0$.
For $g^{<} > 0.14$ the effective potential is repulsive both at
low and high momenta, and attractive in-between as illustrated by
the solid curve in Fig.~\ref{fig4}. Subsequently the energy is minimized for a finite value of $\xi$.

To find the range of parameter space characterized by the appearance of
density-wave states, we need to consider, in addition to stability arguments,
the transition point discussed in Section III. Table 1 summarizes the range of
$s$-wave scattering lengths and critical numbers of $^{52}$Cr atoms
necessary to be inside stable density-wave regime.

\begin{table}[hp]
\renewcommand{\arraystretch}{2.0}
\setlength{\tabcolsep}{3.5mm}
\begin{tabular}{|c|c|c|c|} \hline
\multicolumn{4}{|c|}{\bf TABLE 1} \\
\hline \textbf{$\lambda$} & \textbf{$\ell_z(\mu)$}
& \textbf{$a_{bb}(a_0)$} & \textbf{$N_b(\times 10^{4})$} \\
\hline
\hline
                & 0.15     & 16-21       &  $\geq$ 0.65     \\ \cline{2-4}
0.25            & 0.2    & 16-19       &   $\geq$ 2.3    \\  \cline{2-4}
                & 0.25     & 15-18       &  $\geq$ 5.0 \\
\hline
\end{tabular}
\caption{Tabulation of the scattering length of boson-boson contact interaction
$a_{bb}$ and critical number of bosons $N_b$ for different values of
aspect ratio $\lambda$ and oscillator length $\ell_z$ inside
the density-wave regime
with $N_f=10^3$ and $g_{bf}=70a_0$ for a mixture of chromium isotopes.}
\end{table}

\section{Conclusion}

In summary, we have analyzed the stability of a dipolar bosonic condensate mixed with
non-interacting fermions in a pancake trap at $T=0$. We found that
the fermions help stabilize the condensate for a significant range of boson-boson
and boson-fermion interaction strengths. We then investigated the transition of
the system from a Gaussian-like to the density-wave ground state as a
function of number of bosons, strength of the contact interaction, and
trap aspect ratio. Our central result is the use of a variational ansatz
to show that while in a purely bosonic system the density-wave state is
always unstable it can be stabilized by the admixture of even a small boson-fermion interaction.

In particular, this study leads us to the conclusion that a pancake-shaped
${}^{87}$Rb-${}^{40}$K mixture, which has a large boson-fermion $s$-wave scattering
length, should be absolutely stable in the dipole dominated regime. By tuning the
$s$-wave scattering length it is possible to reach a situation characterized by the
appearance of a roton instability in the excitation spectrum, leading to the existence
of a stable density-wave state. However, due to the small
dipole moment of rubidium atoms the transition to this stable density-wave
regime needs a substantial number of atoms, of the order of $10^6$ .

Future work will discuss the effect of the dipolar nature of
the fermionic isotopes on the condensate -- with chromium atoms in mind --,  the existence and
stability of density waves, as well as possible extensions to rotating systems.

\acknowledgments

We thank Prof. Tilman Pfau for several interesting discussions and his deep insight on dipolar condensates. This work is supported in part by the US Office of Naval Research, by the National Science Foundation, and by the US Army Research Office.

\appendix
\section{Derivation of the energy functional of Eq.~(\ref{enfun})}

In this appendix we derive the energy functional Eq.~(\ref{enfun}) by using
the Gaussian ansatz of Eq.~(\ref{ph}).
First we consider the dipolar interaction energy,
\begin{widetext}
\begin{eqnarray}
E_{\rm dd}=\frac{g_{\rm dd}N_b}{3(2\pi)^2}\int d\bm{k}
\left(3\frac{k^2_z}{k^2}-1 \right)
\exp \left[ -\frac{1}{2} \{ k^2_zd^2_z+(k^2_x+k^2_y)d^2 \} \right ],
\end{eqnarray}
\end{widetext}
where $k^2=k^2_x+k^2_y+k^2_z$. Evaluating this integral gives the form of dipolar energy
\begin{equation}\label{dip1}
E_{\rm dd}=\frac{g_{\rm dd}N_b}{3(2\pi)^2 d^3_z}\left[\frac{2}{3} \eta^2 - F(\eta^{-1})\right],
\end{equation}
with $\eta=d_z/d$ and
$$
F(y)=\frac{\tan^{-1}(\sqrt{y^2-1})}{(y^2-1)^{{3}/{2}}}-\frac{1}{y^2
(y^2-1)}.
$$
Next we calculate the energy due to $s$-wave boson-boson interaction and
fermion-induced interaction. To achieve this goal we break the integral of
the interaction energies expressed in spherical coordinates into two parts,
$E_s \equiv E_s^< + E_s^>$ and $E_{\rm dd}\equiv E_{\rm dd}^< + E_{\rm dd}^>$, according to
\begin{eqnarray}
\int^{\infty}_0 dk
= \int^{2k_f}_0 dk + \int^{\infty}_{2k_f} dk.
\end{eqnarray}

For $k<2k_f$ and in spherical coordinate, the interaction energy including the $s$-wave and induced interaction is given by
\begin{widetext}
$$
E_{s}^{<}+E_{\rm ind}^{<}=\frac{g_{\rm dd}N_b}{2\pi}
\int^{2k_f}_0 \int^{\pi}_0 \left [ g^{<}+
\frac{g^2_{bf} \nu }{4\pi g_{\rm dd}}
\sum_{n=1}^\infty
\left(\! \frac{k}{2k_f} \!\right)^{\!\! 2n}\!\!\!\frac{1}{4n^2-1} \right ]
\exp \left[- \frac{k^2}{2} (d^2_z \cos^2 \theta + d^2 \sin^2 \theta) \right ] k^2dk \sin\theta d\theta
$$
\end{widetext}
where $g^{<}$ is defined by Eq.~(\ref{contact}).
This integral contain contribution from both the local and nonlocal part of the induced interaction. After integrating over $k$ and rearranging the terms we obtain the expression
\begin{equation}\label{in1}
E_s^{<}+E_{\rm ind}^{<}=g^{<}E^{<}_1+g_{\rm ind}  (E^{<}_2 +E^{<}_3),
\end{equation}
with
\begin{eqnarray}\label{ind1}
E^{<}_1(\sigma) &=& \int^{\pi/2}_0 \frac{{\rm erf}(\sqrt{2} k_f \sigma)}{\sigma^3} \sin \theta d\theta, \nonumber\\
E^{<}_2(\sigma) &=& \frac{48 k^3_f}{\sqrt{2 \pi}} \int^{\pi/2}_0
\frac{\exp(-2k^2_f \sigma^{2})}{\sigma^2} \sin \theta d\theta, \nonumber\\
E^{<}_3(\sigma) &=& \frac{3}{\sqrt{2 \pi}} \sum_{n=1}^{\infty}
\frac{2^{n+1/2}}{4n^2-1} \frac{1}{2 k^{2(n-1)}_f} \times \nonumber\\
& & \int^{\pi/2}_0
\left [ \frac{\Gamma(n+3/2) -
\gamma(n+3/2, 2k^2_f \sigma^2)}{\sigma^{2n+3}} \right ] \sin \theta d\theta, \nonumber\\
&&
\end{eqnarray}
Here $\sigma=\sqrt{d^2_z \cos^2\theta+d^2\sin^2\theta}$ is a function of $(\theta,d_z,d)$,
${\rm erf}(y)$, ${\rm erfc}(y)$, $\Gamma$, and $\gamma$ denote
the error function, complementary error function, gamma function,
and incomplete gamma function, respectively.

Similarly, the contribution for $k>2k_f$ to the interaction energy
that contains both the $s$-wave boson-boson interaction and the nonlocal induced interaction is
\begin{widetext}
$$
E_s^{>}+E_{\rm ind}^{>}=\frac{g_{\rm dd}N_n}{2\pi} \int^{\infty}_{2k_f} \int^{\pi}_0
\left [ g^{>} - \frac{g^2_{bf} \nu }{4\pi g_{\rm dd}}
\sum_{n=1}^{\infty} \left( \! \frac{2k_f}{k}\! \right)^{\!\! 2n} \!\!\!\frac{1}{4n^2-1}  \right ]
\exp \left[- \frac{k^2}{2} (d^2_z \cos^2 \theta + d^2 \sin^2 \theta) \right ] k^2dk \sin\theta d\theta,
$$
\end{widetext}
where $g^{>}$ defined in Eq.~(\ref{contact}). In this
equation the first term in the bracket stems from the $s$-wave boson-boson interaction
and the last term from attractive nonlocal part of the induced interaction. Again after evaluating the integral over $k$ we get
\begin{equation}\label{in2}
E_{s}^{>}+E_{\rm dd}^{>}=g^{>}E^{>}_1-g_{\rm ind}E^{>}_2
\end{equation}
where
\begin{eqnarray}
E^{>}_1(\sigma) &=& \int^{\pi/2}_0 \frac{{\rm erfc}(\sqrt{2} k_f \sigma)}{\sigma^3} \sin \theta d\theta, \nonumber
\end{eqnarray}
and
\begin{widetext}
\begin{eqnarray}\label{ind2}
E^{>}_2(\sigma) &=& \frac{3}{\sqrt{2 \pi}} \sum_{n=1}^{\infty} \frac{
{2k_f}^{2(n+1)}}{4n^2-1} \int^{\pi/2}_0 \left [ - \frac{(-1)^{n}
\sqrt{\pi/2}}{(2n-1)!!} \sigma^{2n-3}
{\rm erfc}(\sqrt{2} k_f \sigma) \right . \nonumber\\
& + & \left . \sum_{m=0}^{n-2} \frac{(-1)^m 2^{m+1} (2k_f)^{2m}}{{2k_f}^{2n-2m-3} (2n-3)(2n-5)....(2n-2m-3)}
\exp(-2k^2_f \sigma^{2}) \sigma^{2m} \right ] \sin \theta d\theta,
\end{eqnarray}
\end{widetext}

The total interaction energy is the sum of the contributions in Eq.~(\ref{dip1}), (\ref{in1}), (\ref{in2}) and is given in Eq.~(\ref{enfun}). In this present paper
we took $n=1, \dots, 3$ in the series in Eqs. (\ref{ind1}), (\ref{ind2})
as the energy $E_g$ in Eq.~(\ref{enfun}) converges with this inclusion.


\begin{thebibliography}{99}
\bibitem{nd}
For a fixed number of dipolar bosons, the quantity $n_d$ can be changed by tuning the aspect ratio $\lambda$.
\bibitem{sup}
The qualitative behavior of the total-energy landscape, hence
the stability of the density-wave state, remains unchanged
by the inclusion of the kinetic energy and trapping potential.
\end{thebibliography}
\end{document}